\documentclass{article}

\usepackage[final]{neurips_2024}

\usepackage[utf8]{inputenc} 
\usepackage[T1]{fontenc}    
\usepackage{hyperref}       
\usepackage{url}            
\usepackage{booktabs}       
\usepackage{amsfonts}       
\usepackage{nicefrac}       
\usepackage{microtype}      
\usepackage{xcolor}         
\usepackage{graphicx}
\usepackage{xcolor,hyperref, threeparttable, amsfonts, amsmath, amssymb, natbib}
\usepackage{rotating}

\title{Hierarchical Classification for Predicting Metastasis Using Elastic-Net Regularization on Gene Expression Data}

\author{%
Benjamin Osafo Agyare$^{1}$ \quad Alec Chu $^{2}$ \quad Blessing Oloyede$^3$\\
$^1$Department of Statistics, University of Michigan, Ann Arbor\\
$^2$Department of Cellular and Molecular Pathology, University of Michigan, Ann Arbor\\
$^2$Department of Bioinformatics, University of Michigan, Ann Arbor\\
$^3$Department of Chemical Biology, University of Michigan, Ann Arbor\\
}

\begin{document}

\maketitle

\begin{abstract}
  \textbf{Motivation:} Metastasis is a leading cause of cancer-related mortality and remains challenging to detect during early stages. Accurate identification of cancers likely to metastasize can improve treatment strategies and patient outcomes. This study leverages publicly available gene expression profiles from primary cancers, with and without distal metastasis, to build predictive models. We utilize elastic net regularization within a hierarchical classification framework to predict both the tissue of origin and the metastasis status of primary tumors.\\\\
\textbf{Results:} Our elastic net-based hierarchical classification achieved a tissue-of-origin prediction accuracy of 97\%, and a metastasis prediction accuracy of 90\%. Notably, mitochondrial gene expression exhibited significant negative correlations with metastasis, providing potential biological insights into the underlying mechanisms of cancer progression.\\
\end{abstract}

\section{Introduction}
Cancer metastasis is the spread of cancer cells from a primary tumor site to surrounding tissues or distant locations and contributes to around 90\% of cancer moralities \citep{sewfried2013}. In recent years, cancer incidences have been steadily increasing due to multiple factors such as longer average lifespans and better early detection. To address this, advancement in cancer therapeutics and surgeries has resulted in drastic improvement in prognosis for most localized cancers. However, patient survival continues to be significantly impacted after detection of metastasis, which is most commonly is detected during cancer recurrence and can be years after tumor resection. The appearance of metastasis after tumor resection implies that the majority of cancers have metastasized prior to surgery. This makes detection of cancers with metastatic potential important so that patients can begin conventional metastasis treatments such as surgery, chemotherapy, hormone therapy, immunotherapy, or radiation therapy earlier.

Different cancers also have different metastatic potential. An estimated 6\% of breast cancer patients are presented with metastasis, with bones, brain, liver, and lungs being the most common metastasis location. On the other hand, approximately 60\% of lung cancer patients will have metastasis commonly to the brain, bone, liver, adrenal glands, thoracic cavity, or distal lymph nodes \citep{riihimaki2018}. Differences in metastatic potential and location of metastasis makes early detection and diagnosis paramount for effective treatment. 

For a cancer to metastasize, it must overcome a series of obstacles including detachment from the primary tumor location, intravasate into the circulatory and lymphatic system, evade immune responses and attacks, extravasate at distal capillary locations, invasion of distal locations, and proliferation at distal locations \citep{hunter2008}. In spite of the prevalence of metastasis in cancer patients, the mechanism of metastasis is extremely inefficient and risky for detached cells. Current efforts into using these circulating tumor cells (CTCs) as a biomarkers for metastasis have been inconclusive due to the difficulty of accurately predicting whether these tumor cells are capable of establishing metastasis at distal locations. 

Therefore, new methods and ways to help predict the metastatic potential in patients will be an asset for clinicians. Early detection of metastatic events can drastically improve patient survival by introducing them to therapeutics earlier and treating distal metastasis while it is still small.

In this study, we use publicly available expression datasets to investigate if expression data from primary tissue can be a predictor of metastasis events using a hierarchical classification mechanism as discussed in the next sections of this paper. We then apply the model to examine if there are shared underlying gene expressions that can help predict whether a cancer has metastasized or not.


\section{Approach}
\begin{figure}[h!] 
\centering
\includegraphics[width=0.65\textwidth]{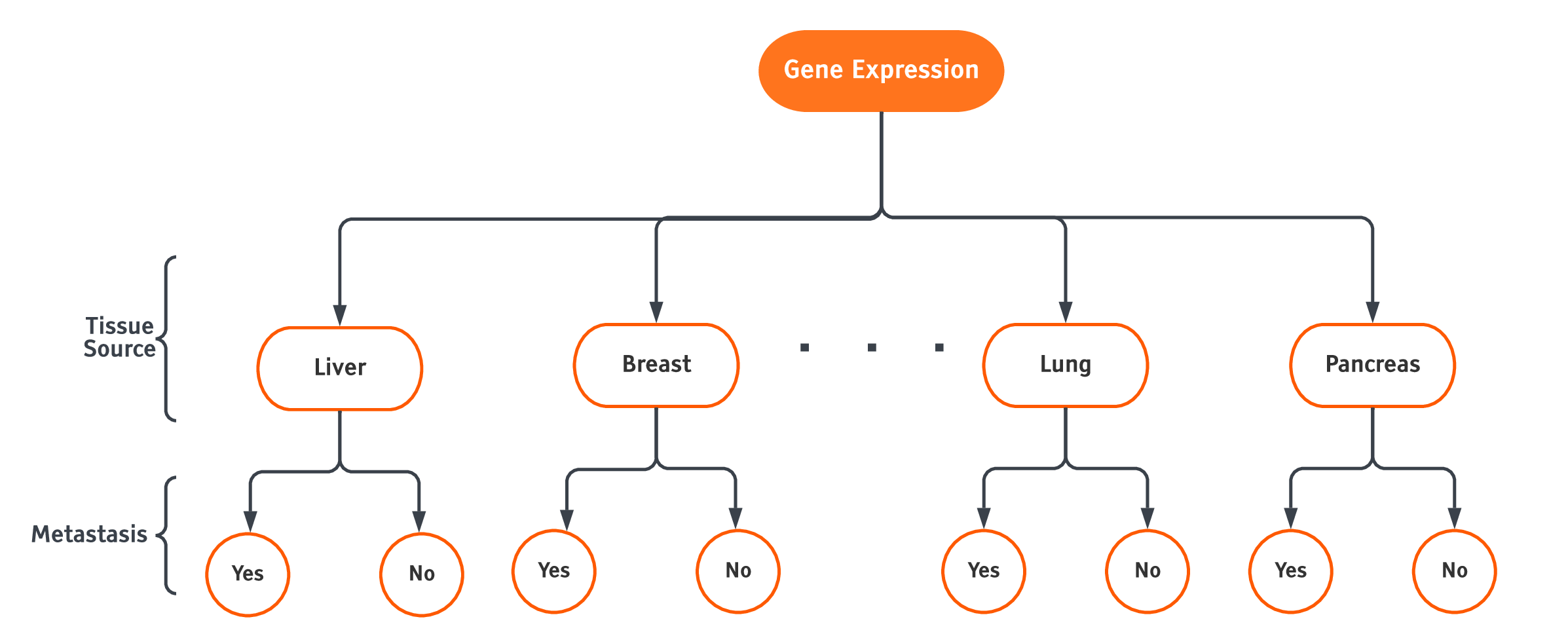}
\caption{Hierarchical structure of the classification task.}\label{fig:hier}
\end{figure}
We implement a hierarchical classification model that follows the structure as seen in Figure \ref{fig:hier}. 

First, train an elastic-net multinomial regression model to classify from which tissue source the tumor grew. Specifically, we predict the tissue source as one of the seven (7) cites, i.e breast,
colon, gastric, kidney, liver, lung, pancreas, and skin. We then fit a hierarchical elastic-net logistic regression model to predict whether the cancer has metastasized. Due to the hierarchical nature of our prediction task, we define accuracy as ability to first correctly predict the tissue source, and then conditionally predicting metastasis given the predicted tissue sources and the gene expressions.

\section{Methods}
\subsection{Datasets}
\subsubsection{Data sources}
Expression data from various primary tissues with and without metastasis were obtained from the Genomic Data Commons Data Portal (GDC). We chose primary tissues of the most common tissue types on the GDC and compiled a total of approximately 5000 samples. These samples represent data collected from various studies including The Cancer Genome Atlas Program (TCGA), Clinical Proteomic Tumor Analysis Consortium (CPTAC), Human Cancer Models Initiative (HCMI), Count Me In (CMI). Around 400 additional datasets were also gathered from various independent studies available on Gene Expression Omnibus (GEO), including \citep{kim2014} \citep{siegel2018} \citep{rothwell2014} \citep{mcdonald2017} \citep{Badal2017} \citep{yang2017} \citep{wang2021}  \citep{menck2022}, to further increase the number of metastasis samples and introduce different types of expression data. Overall, our samples are represented by about 23\% Breast, 11\% Colon, 8\% Gastric, 14\% Kidney, 8\% Liver, 20\% Lung, 2\% Pancreas, and 15\% Skin tissues. With respect to metastasis, approximate 80\% of samples were primary tissues without metastasis while 20\% were from primary tissues with metastasis. 
\subsubsection{Data-Prepossessing}
To simplify the model, we filtered out non-protein-coding genes from the dataset such as lncRNAs and ncRNAs to remove sources of confusions. 
To ensure standardization of the 19,938 features, we utilized transcripts per millions (TPM), then transformed the features by Z scores to obtain unit variance across features. This ensures that differences between samples and methods can be normalized to sequencing depth and that no feature will dominate the predictive power of the model by their raw scale.

As a matter of key relevance, it is noteworthy that we standardize the test feature sets based on the mean and standard deviation of the training features in order to reduce data leakage. The section below discusses in detail the strategies adopted in splitting the data. 

\subsubsection{Data-Splitting}
To assess the performance of the models under consideration, we adopt training-testing splitting where the training set comprises 70\% of the dataset, totalling 3,875 samples while the testing set takes up the remaining 30\%, totalling 1,665 samples. To foster adequate representation of all tissue sources, especially as our data is imbalanced, the split was performed using stratified sampling. 

The training set was further split into cross-validation sets. This step is highly imperative in model selection and hyper-parameter tuning. Specifically, by randomly splitting and assigning each the 3,875 training samples into 10 folds, we obtain a training and validation sample of 3,488 and 387 respectively, yet ensuring proportionate representation of each tissue source across all splits. 

\subsection{Modeling}
\subsubsection{Multinomial Model} 
Given our multi-class prediction task for the tissue source prediction, we use the multinomial model which extends the binomial when the number of classes is more than two \citep{glmnet}. Suppose the response variable has $K$ levels $\mathcal{G}=\{1,2, \ldots, K\}$, and features $X \in \mathcal{R}^{N \times p}$ for a dataset of sample size $N$ with $p$ predictors. Here we model

\begin{equation}\label{eqn:multonom}
  \operatorname{Pr}(G=k \mid X=x)=\frac{e^{\beta_{0 k}+\beta_k^T x}}{\sum_{\ell=1}^K e^{\beta_{0 \ell}+\beta_{\ell}^T x}}  
\end{equation}
This structure induces a linear predictor for each class.

\subsubsection{The Elastic-Net Model} 
The elastic net \citep{zou2005} is a regularized method that coalesces the $L_1$ and $L_2$ penalties of the lasso and ridge regression methods by learning from their shortcomings to improve the regularization of statistical models. It does so by linearly combining the variable selection feature of the lasso and parameter shrinkage property of the ridge model simultaneously. This effective regularization technique allows for controlling multicollinearity, performing regression in high dimensional data settings $(p>>n)$, and for reducing excessive noise in our data to allow for isolating the most influential variables while balancing prediction accuracy \cite{Boehmke2019HandsOnML}. Specifically, for a multi-class prediction task given by \eqref{eqn:multonom}, we specify our model is as follows:

Let $Y$ be the $N \times K$ indicator response matrix, with elements $y_{i \ell}=I\left(g_i=\ell\right)$. Then the elastic net penalized negative log-likelihood function is given by:

\begin{multline}\label{eqn:e-net}
\ell\left(\left\{\beta_{0 k}, \beta_k\right\}_1^K\right)=-\frac{1}{N} \sum_{i=1}^N\left(\sum_{k=1}^K y_{i l}\left(\beta_{0 k}+x_i^T \beta_k\right)\right) \\
+\frac{1}{N}\log \left(\sum_{\ell=1}^K e^{\beta_{0 \ell}+x_i^T \beta_{\ell}}\right) +\lambda\left[(1-\alpha)\|\beta\|_F^2 / 2+\alpha \sum_{j=1}^p\left\|\beta_j\right\|_1\right]
\end{multline}

Here $\beta$ is a $p \times K$ matrix of coefficients. $\beta_k$ refers to the $k$ th column (for outcome category $k$ ), and $\beta_j$ the $j$ th row (vector of $K$ coefficients for variable $j$ ) \cite{glmnet}. The tuning parameters $\lambda \geq 0$ and $\alpha \in [0,1]$ control the amount of regularization, and the mixing rate of the ridge and lasso penalties respectively. Thus, setting $\alpha = 0$ leaves a ridge model while an $\alpha$ value of 1 resets to a lasso model.

\subsection{Tumor cite prediction}
We employed an elastic-net multinomial regression model to predict the tumor source, with the tumor site as the response variable and gene expression levels as the features. The model estimates $\widehat{G}_1 = \operatorname{Pr}(G = k | X)$ for $k \in \{1, 2, \ldots, 7\}$, corresponding to the seven possible tumor sites.

To optimize the model's hyperparameters, we first defined a common fold identifier, ensuring consistent cross-validation folds across all models. We then established a tuning grid for the elastic-net mixing parameter, $\alpha$, ranging from $\{0, 0.1, 0.2, \ldots, 1\}$. A 10-fold cross-validation was performed to determine the optimal penalty parameter, $\lambda \geq 0$, with the best $\lambda$ minimizing the multinomial deviance.

By iterating over the grid of $\alpha$ values, we identified the model that achieved the lowest deviance, which was then selected for subsequent analysis and tumor site predictions.

\subsection{Metastasis prediction}
For metastasis prediction, we followed a similar approach to that used for tumor site prediction. However, given the binary outcome (metastasis: yes or no), we applied an elastic-net logistic regression model. This is a special case of the multinomial regression model when $K = 2$, corresponding to two classes: $\mathcal{G} = \{1, 2\}$.

In alignment with the hierarchical classification framework, our prediction is represented as $\widehat{G}_2 = \operatorname{Pr}(G = k | \widehat{G}_1, X)$, where $\widehat{G}_1$ is the predicted tumor site and $X$ represents the gene expression features. This hierarchical structure allows the model to incorporate the tumor site prediction into the metastasis prediction process.

\section{Results}
\subsection{Tumor cite prediction}
\begin{figure}[tb!] 
\centering
\includegraphics[width=0.6\textwidth]{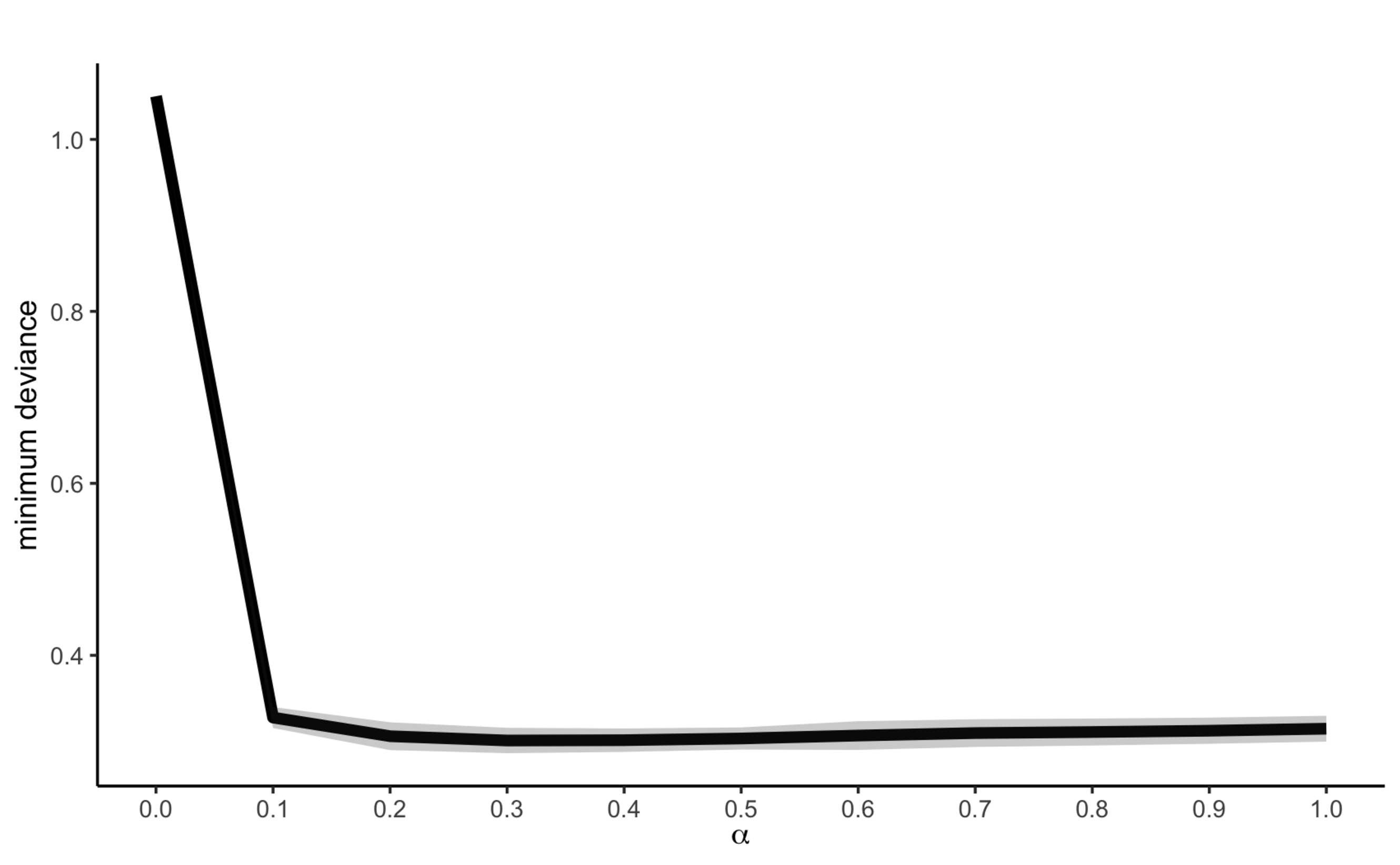}
\caption{Hyperparameter tuning for $\alpha$. The best tuned $\alpha$ is 0.4, as it yields the smallest multinomial deviance. This is used in training our elastic net model for predicting tissue source.}\label{fig:alphas}
\end{figure}

We trained multiple elastic-net models across 10 values of $\alpha$ (ranging from 0 to 1 in increments of 0.1) and used cross-validation to determine the optimal value of the regularization parameter, $\lambda$. The best model was selected based on the value of $\alpha$ that minimized the multinomial deviance. The results from the hyperparameter tuning process are shown in \ref{fig:alphas}.

As depicted in Figure \ref{fig:alphas}, $\alpha = 0$ (which corresponds to the fully ridge model) yields the highest deviance, indicating the poorest performance. The smallest deviance is achieved at $\alpha = 0.4$, which we selected as the optimal value for the elastic-net model. The corresponding $\lambda$ value of 0.007619 was tuned using 10-fold cross-validation.

With these optimal parameters, the elastic-net model achieved a prediction accuracy of 97.36\% for tumor site classification. Given the multi-class nature of the task and the class imbalance, we further assessed model performance using a confusion matrix, as shown in Table  \ref{table:conf_mat_tissue}.

While the overall accuracy is high, the confusion matrix reveals that misclassification rates were higher for the Lung class. Despite this, an accuracy of 97.36\% is notably impressive, particularly when compared to the expected accuracy of approximately 14\% from random guessing across seven classes. These results provide strong confidence in the hierarchical framework used for metastasis prediction, which depends on the accuracy of the tumor site model.

\begin{table}[tb!]
\caption{Confusion Matrix for tissue cite prediction.} 
\label{table:conf_mat_tissue}
\addtolength{\tabcolsep}{-2pt}
\raisebox{-1cm}{\rotatebox{90}{Prediction}}\hspace{3pt}
\begin{tabular}{lcccccccc|c}
\multicolumn{4}{l}{}  & \multicolumn{5}{l}{\hspace{1.1cm}Ground Truth} \\ 
\toprule
 & Breast & Colon & Gastric & Kidney & Liver & Lung & Pancreas & Skin & Total \\ 
\midrule
Breast & 373 &   0 &   0 &   0 &   0 &   1 &   0 &   0  & 374\\ 
  Colon &   1 & 169 &   4 &   1 &   2 &   0 &   2 &   1 & 180\\ 
  Gastric &   1 &   1 & 120 &   1 &   0 &   1 &   1 &   0 & 125 \\ 
  Kidney &   0 &   0 &   0 & 221 &   0 &   0 &   0 &   0 & 221\\ 
  Liver &   0 &   1 &   0 &   0 & 137 &   1 &   0 &   0 & 139 \\ 
  Lung &   2 &   3 &   5 &   2 &   0 & 310 &   5 &   1 & 328\\ 
  Pancreas &   0 &   1 &   2 &   0 &   0 &   2 &  48 &   0 & 53\\ 
  Skin &   0 &   0 &   0 &   0 &   1 &   1 &   0 & 243 & 245 \\ 
   \hline 
Total   &    377 &  175 &   131 &  225 & 140  & 316  & 56 & 245 & 1665\\
\hline

\end{tabular}
\begin{tablenotes}
\item \small The elastic-net at $\alpha = 0.4$ yields an impressive accuracy of 97.36\%. The confusion matrix further details what classes are better predicted. Generally, the model easily misclassifies tissues as lung as seen from the table in terms of prediction error.
\end{tablenotes}
\end{table}

\subsection{Metastasis prediction}
Following a similar procedure of hyperparameter tuning, the logistic regression model for metastasis prediction selected a fully lasso model ($\alpha = 1$) with an optimal $\lambda$ value of 0.0058. Using this model, we achieved a prediction accuracy of 90.33\%. The confusion matrix for metastasis prediction is presented in Table \ref{table:met_result}.

The model demonstrates higher precision in predicting cases where metastasis did not occur, with a precision rate of 91.5\%, compared to 84\% precision for cases where metastasis was present. While the overall accuracy is strong, the difference in precision suggests that the model is more reliable for non-metastasis predictions. Despite this, the results provide a solid foundation for metastasis classification, particularly when combined with the high accuracy of tumor site prediction in the hierarchical framework.

\begin{table}[h!]
\caption{Confusion matrix for metastasis prediction} 
\label{table:met_result}
\addtolength{\tabcolsep}{3pt}
\raisebox{-1cm}{\rotatebox{90}{Prediction}}\hspace{3pt}
\centering
\begin{tabular}{lcc|c}
\multicolumn{1}{l}{}  & \multicolumn{3}{l}{\hspace{1.1cm}Ground Truth} \\ 
  \toprule
 & No & Yes & Total \\ 
  \midrule
No & 1286 & 119 & 1405 \\ 
  Yes &  42 & 218 & 260\\ 
   \hline
Total & 1328 & 337 & 1665 \\
   \hline
\end{tabular}
\end{table}

\subsection{Top Influential genes for metastasis}
\begin{figure}[h!] 
\centering
\includegraphics[width=0.70\textwidth]{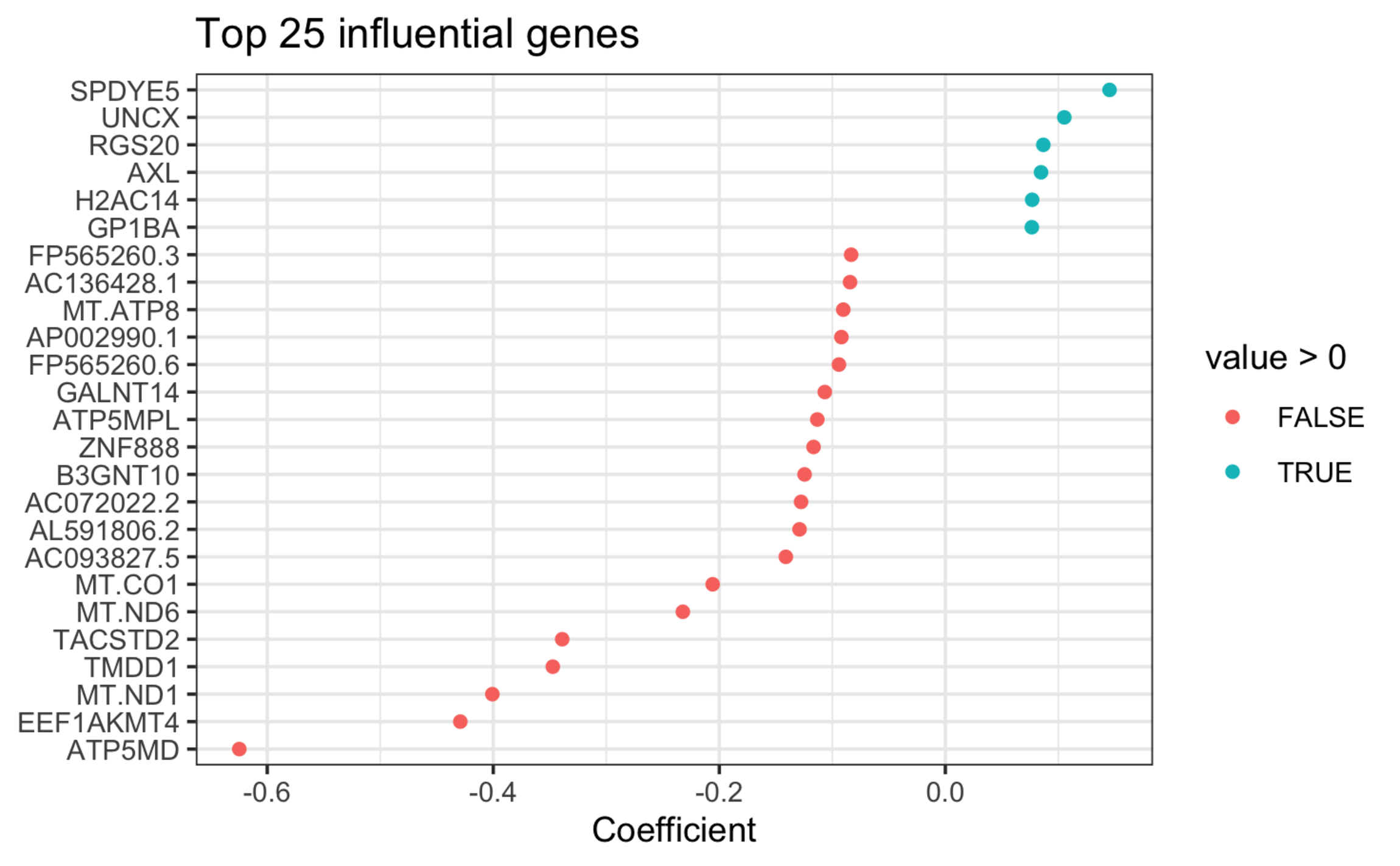}
\caption{Top Influential genes for metastasis prediction}\label{fig:influence}
\end{figure}

To identify the top predictors of metastasis, we analyzed the coefficients of the elastic-net logistic regression model. The top 25 most influential genes were selected based on the absolute values of their coefficients, representing their contribution to the metastasis prediction (Figure \ref{fig:influence}). Of these 25 genes, 6 were positively correlated with metastasis, while 19 were negatively correlated.

Notably, several mitochondrially associated genes involved in the oxidative phosphorylation pathway, such as \textit{ATP5MD}, \textit{MT-ND1}, and \textit{MT-CO1}, were negatively associated with metastasis. This observation aligns with existing knowledge that cancer cells often exhibit altered metabolic processes, and our findings suggest that cancers with a higher metastatic potential may reduce their reliance on oxidative phosphorylation.

Among the positively associated genes, \textit{AXL} emerged as a key gene linked to metastasis. \textit{AXL}, part of the Gas6/AXL signaling pathway, is known to play a role in cancer invasion and metastasis, further reinforcing the biological relevance of our model’s results.

These findings suggest that the genes identified by our model as influential in predicting metastasis align with established literature, indicating that the model’s predictions are biologically interpretable and not driven by artifacts.

\subsection{Overall prediction accuracy}
To evaluate the overall performance of our hierarchical classification algorithm, we assess the ability to predict metastasis given the predicted tissue source. The hierarchical structure of the model introduces varying levels of complexity in assessing prediction accuracy. We categorize the prediction outcomes into three distinct cases:
\begin{enumerate}
    \item Accurately predict metastasis given an erroneously predicted tissue source and vice-versa which we code as \textbf{\textit{semi-accurate}}.
    
    \item Accurately predicting metastasis given accurate prediction of the tissue source, coded as \textbf{\textit{accurate}}, and

    \item Inaccurately predicting metastasis given an erroneously predicted tissue source, coded as \textbf{\textit{inaccurate}}.
\end{enumerate}
Based on this classification scheme, our hierarchical algorithm achieves an overall prediction accuracy of 97\% for the \textbf{\textit{accurate}} case, 3\% for the \textbf{\textit{inaccurate}} case, and 0\% for the \textbf{\textit{semi-accurate}} case. These results highlight the robustness and reliability of our hierarchical classification framework, with the model consistently delivering highly accurate predictions.

%
%

\section{Discussion}

The expression profiles of cancer can be very distinct between tissue of origin as well as between individuals, which makes establishing general trends among different cancers difficult. Our model was able to achieve an accuracy of 97\% in predicting tissue of origin and a 90\% for predicting whether the tumor has metastasis given the tissue of origin.

It is know that recurrent oncogene mutations are often used as a biomarker in cancer classification. What is unique about our study is the usage of expression profiles without including gene mutations. The complexity of annotating and understanding the effects of different gene mutations makes developing models based on gene mutations difficult for all but the most recurrent mutations. However, by using expression data regardless of mutation status, we provide a model that can be more easily understood and representative of the cell biology of cancer cells.

However, there are many confounders which limits the accuracy of the model. Although we tried to include as many samples as possible across different studies, batch effects within studies may result in differences between samples that were collected as primary and samples that were collected for metastasis studies. Additionally, patients for whom both primary and metastasis samples are sequenced may not represent the whole population of patients with metastasis, as primary tissues of patients with metastasis are typically late-stage cancers while normal primary tissues may be gathered from any stage. These effects may result in those primary samples with metastasis representing severity and development of cancers rather than metastasis potential.

Overall, the results are promising in showing that there may be sufficient evidence in expression profiles of primary tumors that can predict metastasis events. Further studies and incorporation of additional datasets may help with improving the accuracy of the model. 

\vspace{0.2in}

\section*{Funding}
This project was not funded by anyone in particular.
\bibliographystyle{natbib}
\bibliography{document}

\begin{thebibliography}{}

\bibitem[Badal~B(2017)Badal~B]{Badal2017}
Badal~B, Solovyov~A, D. C. S. C. J. e.~a. (2017).
\newblock Transcriptional dissection of melanoma identifies a high-risk subtype underlying tp53 family genes and epigenome deregulation.
\newblock {\em The Journal of Clinical Investigation\/}, {\bf 2(9)}.

\bibitem[Boehmke and Greenwell(2019)Boehmke and Greenwell]{Boehmke2019HandsOnML}
Boehmke, B.~C. and Greenwell, B.~M. (2019).
\newblock Hands-on machine learning with r.

\bibitem[Hastie~T(2021)Hastie~T]{glmnet}
Hastie~T, Qian, J. T.~K. (2021).
\newblock An introduction to glmnet.

\bibitem[Hunter(2008)Hunter]{hunter2008}
Hunter, K.~W., C. N. P. . A.~J. (2008).
\newblock Mechanisms of metastasis.
\newblock {\em Breast cancer research: BCR\/}, {\bf 10}, S1.

\bibitem[Kim~SK(2014)Kim~SK]{kim2014}
Kim~SK, Kim~SY, K. J. R. S. e.~a. (2014).
\newblock A nineteen gene-based risk score classifier predicts prognosis of colorectal cancer patients.
\newblock {\em Molecular Oncology\/}, {\bf 8}, 1653–1666.

\bibitem[McDonald~OG(2017a)McDonald~OG]{mcdonald2017}
McDonald~OG, Li~X, S. T. T. R. e.~a. (2017a).
\newblock Epigenomic reprogramming during pancreatic cancer progression links anabolic glucose metabolism to distant metastasis.
\newblock {\em Nature Genetics\/}, {\bf 49(3)}, 367–376.

\bibitem[McDonald~OG(2017b)McDonald~OG]{yang2017}
McDonald~OG, Li~X, S. T. T. R. e.~a. (2017b).
\newblock Recurrently deregulated lncrnas in hepatocellular carcinoma.
\newblock {\em Nature Communications\/}, {\bf 8}, 14421.

\bibitem[Menck~K(2022)Menck~K]{menck2022}
Menck~K, Wlochowitz~D, W. A. C. L. W. A. S. A. K. U. W. S. S. H. B. H. W. E. P. T. H. K. B. T. B.~A. (2022).
\newblock High-throughput profiling of colorectal cancer liver metastases reveals intra- and inter-patient heterogeneity in the egfr and wnt pathways associated with clinical outcome.
\newblock {\em Cancers (Basel)\/}, {\bf 14(9)}, 2084.

\bibitem[Riihimäki(2018)Riihimäki]{riihimaki2018}
Riihimäki, M., T. H. S. K. S. J. . H.~K. (2018).
\newblock Clinical landscape of cancer metastases.
\newblock {\em Cancer medicine\/}, {\bf 7(11)}, 5534--5542.

\bibitem[Rothwell~DG(2014)Rothwell~DG]{rothwell2014}
Rothwell~DG, Li~Y, A. M. T. C. e.~a. (2014).
\newblock Evaluation and validation of a robust single cell rna-amplification protocol through transcriptional profiling of enriched lung cancer initiating cells.
\newblock {\em BMC Genomics\/}, {\bf 15(1)}, 1129.

\bibitem[Seyfried(2013)Seyfried]{sewfried2013}
Seyfried, T.~N., . H. L.~C. (2013).
\newblock On the origin of cancer metastasis.
\newblock {\em Critical reviews in oncogenesis\/}, {\bf 18(1-2)}, 43--73.

\bibitem[Siegel~MB(2018)Siegel~MB]{siegel2018}
Siegel~MB, He~X, H. K. H. A. e.~a. (2018).
\newblock Integrated rna and dna sequencing reveals early drivers of metastatic breast cancer.
\newblock {\em The Journal of Clinical Investigation\/}, {\bf 128(4)}, 1371–1383.

\bibitem[Wang(2021)Wang]{wang2021}
Wang, B., Z. Y. Q. T. e.~a. (2021).
\newblock Comprehensive analysis of metastatic gastric cancer tumour cells using single-cell rna-seq.
\newblock {\em Sci Rep\/}, {\bf 11}, 1141.

\bibitem[Zou(2005)Zou]{zou2005}
Zou, H., . H.~T. (2005).
\newblock Regularization and variable selection via the elastic net.
\newblock {\em Journal of the Royal Statistical Society. Series B (Statistical Methodology\/}, {\bf 67(2)}, 301--320.

\end{thebibliography}

\end{document}